\documentstyle[11pt,aaspp4]{article}
\def\mic{$\mu{\rm m}$ }
\received{13 August 1997}

\lefthead{Rigaut et al.}
\righthead{CFHT AOB Performance}

\begin{document}

%
%

\title{Performance of the Canada--France--Hawaii Telescope
Adaptive Optics Bonnette}

\author{F. Rigaut\altaffilmark{1}, D. Salmon, R. Arsenault, J. Thomas}
\affil{Canada--France--Hawaii Telescope, 65--1238 Mamalahoa Hwy, Kamuela
HI-96743, USA\\ Email: {\tt frigaut@eso.org, {\sl name}@cfht.hawaii.edu}}

\author{O. Lai, D. Rouan$^2$, J.P. V\'eran$^{2,3}$, P. Gigan}
\affil{Observatoire de Paris, Section de Meudon, 92190 Meudon principal Cedex, France\\ Email : {\tt lai@hplyot.obspm.fr, daniel@isolde.obspm.fr, veran@hplyot.obspm.fr,
pierre.gigan@obspm.fr}}

\author{David Crampton$^2$, J.M. Fletcher, J. Stilburn}
\affil{Dominion Astrophysical Observatory, HIA, NRC, Victoria, V8X 4M6, Canada\\ Email : {\tt {\sl First.Lastname}@hia.nrc.ca} }

\author{C. Boyer and P. Jagourel}
\affil{CILAS, route de Nozay, 91460 Marcoussis,
France}

\altaffiltext{1} {Now with the European Southern Observatory, Karl
Schwarzschild Str.2, D-85748 Garching b. M\"unchen}
\altaffiltext{2}{Visiting Astronomers, Canada-France-Hawaii Telescope,
which is operated by the National Research Council of Canada, the
Centre National de la Recherche Scientifique, and the University of
Hawaii}
\altaffiltext{3} {Also with the Ecole Sup\'erieure
des T\'el\'ecommunications, 46, rue Barrault, F-75634 Paris Cedex 13, France}

\begin{abstract}
Extensive results from the commissioning phase of PUEO, the adaptive
optics instrument adaptor for the Canada--France--Hawaii telescope
(CFHT), are presented and discussed. Analyses of more than 750 images
recorded with a CCD and a near-IR camera on 16 nights in wavelengths
from $B$ to $H$ are used to derive the properties of the compensated
wavefront and images in a variety of conditions. The performance
characteristics of the system are analyzed and presented in several
ways, in terms of delivered Strehl ratios, full-width-half-maxima
(FWHM), and quantities describing the improvements of both. A
qualitative description is given of how the properties of the
corrected images result from the structure function of the
compensated phase. Under median seeing conditions, PUEO delivers
essentially diffraction--limited images at $H$ and $K$, images with
FWHM$\sim$0\farcs1 at $J$ and $I$, and provides significant gains down
to $B$, with guide stars as faint as $R$ = 14. During good conditions,
substantial gains were realized with guide stars as faint as $R$ =
17. A simple user-interface and software which automatically and
continuously optimizes the mode gains during observations makes the
operational efficiency extremely high.  A few astronomical examples
are briefly discussed.

\end{abstract}


\section{INTRODUCTION}

The Adaptive Optics Bonnette (AOB), also called PUEO after the sharp
eyed Hawaiian owl, was developed for the Canada-France-Hawaii
Telescope, based on F. Roddier's curvature concept (Roddier et al.
1991, Arsenault et al. 1994).  The ``bonnette'' (adaptor) is a facility
instrument mounted at the f/8 Cassegrain focus of the CFH 3.6m
telescope on top of Mauna Kea, Hawaii.  The instrument is the result of
a collaborative effort between several institutes: CFH managed the
project and designed and produced the general user interface. The
Dominion Astrophysical Observatory (DAO, Canada) designed and
fabricated the opto-mechanical bench, the curvature wavefront sensor
(WFS) (Arsenault et al. 1994, Graves \& McKenna 1991, Graves et al. 1994)
and the electronics for both. The French company CILAS (formerly
Laserdot) provided the deformable mirror (DM) and the real time
computer hardware and software, including a high level maintenance
interface. The Observatoire de Paris--Meudon manufactured the separate
tip--tilt mirror and was in charge of the final integration, testing
and calibration of the entire instrument. The UH adaptive optics team
provided guidance throughout the project. The system was commissioned
at CFHT during three runs in the first semester of 1996. In this paper,
we first briefly describe the instrument and then present its
performance, mostly in term of image improvement. The properties of the
compensated images are also discussed.

\section{INSTRUMENT DESCRIPTION}

PUEO has relatively few optical components, mostly reflecting  ones
(see Fig. 1).  The telescope beam  is initially
diverted by a flat mirror into PUEO. This mirror can be withdrawn,
allowing the beam to pass straight through to a focus which is confocal to
that of PUEO, allowing rapid switchovers from the f/19.6 corrected beam
to the direct f/8 beam if desired.  The optical design (Richardson
1994) includes an f/8 off-axis parabola that collimates the beam and
images the telescope pupil on the 19 electrode deformable mirror.  A
f/19.6 off-axis parabolic mirror, mounted on a fast tip-tilt platform, then
directs the beam to the science instrument.  The use of off-axis
parabolic mirrors allows the essential optical requirements to be
achieved in a compact instrument with small optical components, helping
to minimize flexure. The flexure of the optical components in the
science path amounts to less than a few microns per hour of motion on
the sky.  Optionally, an atmospheric dispersion compensator (ADC) can
be inserted in the collimated beam for observations at visible
wavelength up to zenith distances of 60 degrees. In addition,  a target
and pinhole sources can be inserted at the location of the telescope
focus for calibration and testing.

Prior to the science focus, a beamsplitter reflects part of the light
to the visible wavefront curvature sensor. This wavefront sensor is
mounted on a three-axis stage which can be remotely controlled to allow
selection of a reference star different from the science object at any
location within the 90\arcsec\ diameter field. Unfortunately, at least
one of the axes of the WFS developed backlash during integration and
testing so that the flexure between the WFS and the science focus
exceeds the original specifications and now can be up to $\sim$10\mic
per hour, depending on the orientation of the system.

The DM and WFS geometry (19 electrodes and subapertures divided up into two
rings plus a central electrode) is well suited to circular pupils; the
inner ring and the central electrode allow solution of Poisson's
equation over the pupil while the outer ring provides the required boundary
conditions (Roddier 1988, Rousset 1994).  Such a system, with relatively few
degrees of freedom but a high bandwidth, is particularly well suited to
Mauna Kea seeing conditions where turbulence is weak yet fast (Racine
et al. 1991, Roddier et al. 1990).

Modal control and mode gain optimization (Gendron \& L\'ena 1994,
Ellerbroek et al. 1994, Rigaut et al. 1994) maximize the instrument
performance according to the state of turbulence and the guide star
magnitude. We have modified the modal control as presented in Gendron
\& L\'ena (1994) to adapt it to closed--loop operation. Using power
spectra of the mode coefficients from about the previous 30s,
combined with a model of the closed--loop transfer functions
(calibrated in the laboratory), new mode gains are computed and updated on
a time scale between 30s and 2 minutes, allowing the system to track
slow seeing variations. In practice this works extremely well.

PUEO was extensively tested in the laboratory at 0$^{\rm o}$ and 20$^{\rm o}$ C 
for flexures, optical quality, delivered performance and bandwidth (Lai et al. 1996). The main characteristics and results are summarized in
Table 1. 

The quoted $\lambda/20$ rms at 500 nm in Table 1 refers to the optical
quality of the science path from the input focus to the output focus,
after the mirror shape has been adjusted using an interferometer as a
wavefront sensor located at the science focus. 
When the curvature WFS is used to cancel the optical aberrations of
AOB, the optical quality of the science path is $\lambda/8$. The major
part of this degradation (from $\lambda/20$ to $\lambda/8$) most likely comes from non--common path errors: optical aberrations of
the WFS optics are measured and compensated, and therefore are erroneously
introduced in the science path (in PUEO, there is no easy way to
calibrate out these WFS aberrations). Similarly, optical aberrations of the
science path after the beamsplitter are not measured, and therefore stay
uncompensated. Part of the error may also come from cross--talk and
aliasing effects, a consequence of the limited number of degree of
freedom of the WFS, but we have no quantitative estimates of the
magnitude of these effects.

During the observing runs reported in this paper, we used focal
enlargers (both in the visible and in the near IR to adapt the CCD/IR
array sampling) which further degraded the image quality down to
approximately $\lambda/6$ rms at 500nm. This is still acceptable --
although marginally -- compared to the residual $\lambda/2$ rms at 500
nm typical of compensated images (see section~\ref{performance_on_the_sky}).

\section{PERFORMANCE ON THE SKY}
\label{performance_on_the_sky}

PUEO was extensively tested on the telescope during three observing
runs in the first semester of 1996. The performance was evaluated in
both the visible and the near infrared wavebands, using a 2K$\times$2K
CCD and a 256x256 NICMOS array (loaned by the Universit\'e de
Montr\'eal).  The emphasis was put on the performance in the near IR,
where the instrument is expected to realize its full potential. The
commissioning tests included purely engineering tests, performance
evaluation tests and scientific programs. The latter were intended to
test in ``real life" what could be achieved by the instrument, and to
set up the data acquisition and reduction procedures. The engineering
tests were to check that all the functions performed as expected and
make the necessary calibrations (wavefront sensor motions, ADC
calibrations, etc.).  In this section, we report on the result of the
performance evaluation, in terms of wavefront and image
characteristics.

\subsection{Operational Efficiency}
\label{system_behavior}

One of the main goals in designing this system was to make a
user--friendly, robust interface (Thomas et al. 1997).  The user is
presented with a limited choice, simple interface: basically
one-button ``Start/Stop compensation''.  This turned out to be
achievable, and efficient both in term of system operation and
performance. It covers all cases, from the brightest to the dimmest
objects (${\rm m}_{\rm R}=17$), thanks to the closed--loop optimized
modal control.  In turn, the operational efficiency of the system is
very high:  the set-up on an object, including initiation of the procedure to
optimize the modal gains, takes less than one minute. In fact, because the
instrument focus is taken care of by the adaptive compensation, the
overhead is actually less than for a standard imager.

\subsection{Turbulence characterization and PSF files}
\label{turbulence_characterization}

Directly associated with the modal control, and in addition to the
capabilities described above,  we have implemented a tool that allows
{\sl a posteriori} determination  of the system state during a science
exposure (V\'eran et al. 1997). The covariance of the wavefront sensor
measurements, the deformable mirror command covariance, and other
parameters are computed simultaneously during each exposure, and stored
in ``{\sl PSF files}".  All kinds of diagnostics can be made from these
data and, in particular, we can compute the $D/r_0$ (where D is the
telescope diameter and  $r_0$ is the Fried parameter) and the system
PSF (point-spread-function) for a particular exposure.   A detailed
discussion of the method used to retrieve the PSF from the system data
is given in V\'eran et al. (1997).  The usefulness of having an estimate of
the PSF simultaneous  with the science exposure versus having to image
a PSF calibration star before, during, and/or after cannot be
overstressed. In addition to the gain in efficiency, having an estimate
of the PSF under precisely the same atmospheric conditions as during
the science exposure (by definition) is essential for many astronomical
projects.

The $D/r_0$ values derived from the ``PSF files" are used in all the
following discussions.  $D/r_0$ was computed by fitting the actual
variance of the system modes, corrected for noise and spatial aliasing
(excluding tip-tilt) with their theoretical Kolmogorov value. The
statistical contribution of noise is computed from the number of
photons available for the measurements, and the effects of aliasing are
estimated from a calibrated numerical model of the WFS. A detailed
discussion of this method can be found in V\'eran et al. (1997).  Each
$r_0$ determination was made over a period of time of at least 30
seconds.  The distribution of $D/r_0$ (540 measures in 16 nights spread
over a period from March to October 1996) is shown in
Figure 2.  It is well fitted by a log--normal
distribution with a mean $r_0 = $15.5 cm at 500 nm.  This corresponds
to a seeing disk of 0\farcs67.  This determination includes both free
atmosphere seeing, dome seeing and mirror seeing (although the latter
have short, to very short, outer scales and may not be accounted for
properly). Because tip-tilt is excluded in the $D/r_0$ calculation,
telescope jitter or free atmosphere outer scale effects are not
included. It can be noted that telescope jitter is not expected to be
of a very large amplitude on telescopes like CFHT, because they are
massive and relatively well protected from the wind. No conspicuous
peak has never been noticed in the power spectra of the tip-tilt
components. The $D/r_0$ values used were computed for the actual
direction of observation and were {\sl not} corrected to the zenith. If
we assume an average zenith distance of 30\arcdeg, the median seeing at
zenith becomes 0\farcs58.  This compares well with values derived from
other data sets for the same Mauna Kea site (Roddier et al. 1990,
Racine et al.  1991).  The calculated $D/r_0$ values were also checked
against values determined from open--loop exposures. The error on
$D/r_0$ is a few percent (2-5\%). Something worth noting is that most
of the time the atmosphere exhibited a very good match with a
Kolmogorov type turbulence. On some occasions (10 to 20\% of the time),
we noted deviations that may be attributable to dome or mirror seeing.

\subsection{System Performance: Wavefronts}
\label{system_performance:wf}

Of fundamental importance in understanding the performance and
efficiency of an AO system is its actual ability to compensate
wavefront distortions. One way to investigate this problem is by using
a modal decomposition of the phase. A natural set of modes is the
mirror modes, which are used in our modal control. These mirror modes
have been generated from the deformable mirror influence functions, and
orthogonalized using a Graam-Schmitt process to insure the best
possible correspondence to the Zernike modes for reasons of convenience
(in off--loading the low  frequency temporal variations of tip, tilt and
defocus to the telescope control system), and because the statistical
properties of the Zernike modes in Kolmogorov turbulence are well
known.  There are 19 mirror modes and the first 15 are very similar to
the first 15 Zernike modes. The remaining 4 are similar to Z$_{18}$,
Z$_{20}$, Z$_{21}$, Z$_{28}$.

Figure 3 shows an example of such a modal
decomposition.  The variance of the mirror mode coefficients is plotted
against the mode number, for the input and output wavefronts (before
and after compensation).  The variance of the mode coefficients for the
input wavefront was computed from the deformable mirror commands. For
the output wavefront, it has been computed from the WFS measurements.
In both cases, the {\sl contribution of noise and spatial aliasing has
been taken out} (V\'eran et al. 1997), so that these represent the
actual contents of each mode in the wavefront phase, unlike what has
been reported in most previous work.

This figure shows an excellent match between the actual coefficient
variances in the uncompensated phase (solid line) and the Kolmogorov
fit for our telescope (diamonds), that is, including the central obstruction.
This is typical of the behavior observed
throughout our runs, except in a few cases, as noted earlier, which
show deviations attributed to mirror or dome seeing. The attenuation of
the tip and tilt modes is due either to the fact that we off--load the
very low temporal frequency tip--tilt fluctuations to the telescope
control sysyem (this is therefore not taken into account in these
calculations) or to a finite outer scale of turbulence. If completely
attributed to an outer scale effect, the work of Winker (1991) leads to
$L_0=77$m, again a typical value for our data set. Because it is
difficult to disentangle effects of tracking from outer scale effects,
this value is tentative. It can, however, be
taken as a lower limit for $L_0$.

Figure 3 shows that the compensation attenuates the
coefficient variance by a factor which varies from approximately 30
(for tip--tilt) down to almost no attenuation for the highest modes.

\subsection{System Performance: Images}
\label{system_performance:images}

All images (IR and visible) were reduced following standard image
reduction procedures. In order to get statistically meaningful data,
Strehl ratios (ratio of the maximum of the actual image to the maximum
of the theoretical diffraction--limited image) and FWHM values were extracted
only from long exposures, with integration times of $>$15s. Only a few
uncompensated images were taken, basically to calibrate our $D/r_0$
estimator. {\sl Strehl ratios reported throughout this paper have been
corrected for the static aberrations of PUEO and the camera system},
i.e., the actual images were compared not to fully diffraction--limited
images, but to images obtained using the artificial point
source with no turbulence.  The Strehl ratio of the latter ``static"
images are reported in Table 2. As noted above, they are
equivalent to an optical quality of $\lambda/5$ rms at
500 nm for the IR camera and $\lambda/7$ rms at 500 nm for the visible
camera.


Figures 4 and 5 present the performance
of PUEO in terms of the Strehl ratio of the compensated images. Figure 4
shows the histogram of Strehl
ratios obtained in the $J$, $H$ and $K$ bands.
In this plot, as well as in Fig. 6, we have 
included only those Strehl ratios derived from images of ``bright" stars, with
R $<$13.5 mag. 
Figure 5 is a plot of the Strehl ratio versus the Fried
parameter $r_0$ at the image wavelength.  Note that $r_0$ varies as
$\lambda^{1.2}$, so that the median value $r_0 = 15.5$ cm at 500 nm
translates into a median value of 46 cm at 1.25 \mic ($J$ band), 65 cm
at 1.65 \mic ($H$) and 93 cm at 2.23 \mic ($K$) (see also
table 2). Again, note that these values are not
given for the zenith, but derived from observations at the actual target position.  The lower solid line is the Strehl ratio of the seeing--limited image.
The points exhibit little scatter, mostly thanks to the simultaneous
estimates of $r_0$, as discussed in
section~\ref{turbulence_characterization} and to the fact that the
system bandwidth on bright guide stars is, in most cases, several times
larger than the Greenwood frequency.  The points agree well with the
theoretical curve (dashed line) which corresponds to full compensation
of eight Zernike modes (or a compensated phase variance equal to 0.052
$\times (D/r_0)^{5/3}$). The dotted line represents the simulated
performance, expected if only fitting and aliasing errors are contributing.
This corresponds to a compensated phase variance of 0.041 $\times
(D/r_0)^{5/3}$, equivalent to approximately ten fully--corrected
Zernike terms (Rigaut et al. 1994, V\'eran et al. 1997).  The
difference between the number of modes controllable by the system (19)
and this figure of ten comes from (a) the piston, which is not
measurable, has no effect on image quality and is therefore not
corrected, (b) four of our modes which are high order and therefore not
very efficient in term of phase variance reduction, and, most
importantly, (c) spatial aliasing, a feature intrinsic to any system, often
underestimated.

The difference between ten modes (as expected when only fitting and aliasing errors are considered) and eight modes (achieved) is due to: 
\begin{enumerate}
\item Noise, always non-negligible when using photon-counting devices
such as APDs. For a R = 11.6 mag  guide star, the average noise
contribution to the phase error is approximately 1.0 radian$^2$ at 0.5
\mic , which corresponds to 0.006 $\times (D/r_0)^{5/3}$ for median
seeing. This accounts for the main bulk of the difference between the
0.041 and 0.052 $\times (D/r_0)^{5/3}$ mentioned above.
\item The finite temporal bandwidth. As already mentioned, the temporal
error is usually very small for bright guide stars (see the
discussion related to figure 9 below for the behavior
of the temporal error on fainter guide stars),
but on some occasions, when the turbulence
is exceptionally fast -- jet stream flowing above Mauna Kea --, it can
contribute to the global error budget. Unfortunately, we have no way to
measure this on a routine basis on our system (we can record sequences
of measurements and mirror commands, but not simultaneously with an
image, and not on a routine basis, given the amount of data space
required). 
\end{enumerate}

Another small contribution comes from the uncorrected part of the
primary mirror figure. This has been measured to amount to approximately
to 0.06 \mic rms, or $\lambda/10$ rms at 0.5\mic, a value in accordance
with direct measurements of the CFH telescope optical quality
(C.Roddier, private communication).  Note that this performance of
eight fully--compensated Zernike modes for 19 actuators is comparable to,
although slightly better than, that derived for the Come-On system
(Rigaut et al. 1990), an earlier version of the current Adonis (Beuzit
et al. 1995) which also had 19 actuators.

The ratio of the achieved Strehl ratio (corrected for the static
aberrations of the optical bench) to the uncompensated image Strehl
ratio (a theoretical expression, therefore also assuming no further
degradation by optical aberrations, telescope jitter, etc.,), gives the
gain in peak intensity, shown in Figure 6.  Median gains
for each bandpass are also listed in Table 2.  These
gains translate directly into substantial gains in sensitivities for
unresolved sources (up to 2.5 mag in $J$ and $H$). The upper
solid line in Figure 6 is the limit imposed by 
diffraction: if the image is fully diffraction--limited, the Strehl
ratio improvement is equal to one over the Strehl ratio of the
seeing--limited image. The $D/r_0$ values for median seeing conditions
are shown at the bottom of the figure for the $B$ to $K$ bands.

Note that the improvement in the Strehl ratio peaks at  $r_0$ $\sim$50
cm ($D/r_0 = 7$).  This may be compared to the characteristic
length associated with the correction resulting from the geometry of
the system: the DM has 19 electrodes and the average distance $d$
between two electrodes corresponds to $\sim$90 cm, and therefore
$D/d=4$.

In terms of FWHM, the images are basically diffraction limited in $H$
and $K$ for median seeing conditions. A FWHM of $\sim$0\farcs1 is
maintained down to the $I$ band under the seeing conditions usually
encountered at CFHT. The FWHM in the visible region ($B$, $V$ and $R$)
still show substantial gains with respect to the FWHM of uncompensated images
(see below).  Properties of the corrected images, in terms of
morphology, are discussed in the next section.

Figure 7 presents the FWHM improvement achieved by the
PUEO compensation.  The gain is defined as the ratio of the FWHM of the
seeing--limited image, $\lambda/r_0$, to the FWHM of the compensated
image. In the figure, the solid line is the maximum theoretical gain
set by the diffraction limit ($(\lambda/r_0)/(\lambda/D) = D/r_0$)
against $D/r_0$.  Not to involve too many parameters in the analysis,
the improvements in FWHM are shown only for stars brighter than ${\rm
m}_{\rm R} = 13.5$.

The maximum gain in FWHM is obtained for $r_0$ $\sim$40 cm ($D/r_0 =
9$). It is worth noting that the maximum gain in Strehl takes place at
a slightly larger $r_0$ value (50 cm or $D/r_0 = 7$), as mentioned
above.  The $r_0$ value at which the maximum resolution in term of FWHM
is obtained is called the critical $r_0$ value (Rousset et al. 1990),
which, for median seeing conditions, corresponds to a wavelength
that is called the ``critical wavelength" of the system.  For PUEO, this
critical wavelength is 1 micron. Referring to figure 5,
one can see that the biggest gain in resolution is obtained for images
with Strehl ratios between 10 and 15\%.  At $r_0 = 20$ cm ($D/r_0 = 18$),  a
typical value of $r_0$ in the $V$ band, the gain in FWHM is still 2--3.

A different way to present the same FWHM results, which helps forecast
performance during an observing run, is shown in
Figure 8.  The normalized FWHM of the
compensated images is plotted against the $D/r_0$ at the image
wavelength, again for stars brighter than ${\rm m}_{\rm R} = 13.5$.
The normalized FWHM is the FWHM of the image in units of $\lambda/D$ at
the image wavelength. Consequently, it has a lower limit of 1. In the plot,
the upper solid line is the FWHM of the seeing--limited image.  This
figure demonstrates that the normalized FWHM, like the Strehl ratio, is
a characteristic of the system, dependent only upon $D/r_0$ and not
upon the image wavelength.  Two regimes, with a very clear cut-off, are
revealed in this plot.  The first, up to $D/r_0 \approx 8$, is
characterized by high Strehl ratios ($>$20\%) and diffraction--limited
images in terms of FWHM (normalized FWHM $\approx$ 1). Above $D/r_0 = 8$
lies a regime of more partial correction, with low Strehl ratio images
($<$20\%) and FWHM strongly dependent upon the turbulence conditions.
However, as reported above, the gain in FWHM can still be quite
attractive in this domain. This is particularly true for direct
imaging. In the visible, compensated images with FWHM 0\farcs1-0\farcs2 
are commonly obtained with PUEO at CFHT. Even these modest resolution
gains can make a huge difference to the feasibility and efficiency of many
astronomical programs.

Another key issue is how the performance degrades with  guide star
magnitude.  This is shown in Figure 9, in terms of Strehl
ratio attenuation versus the guide star $R$ magnitude. The Strehl ratio
attenuation, ${\cal S}_{\rm att}$, is merely the attenuation with respect
to Strehl ratio values obtained on bright guide stars, under the same
turbulence conditions. This curve was computed using the results
presented in Figure 5. Strehl ratios at $H$ on
faint guide stars were divided by the expected Strehl value for bright
guide stars under the given $D/r_0$ conditions, binned by magnitude and
plotted against the $R$ magnitude of the guide star.  These points were
then fitted using a function of the form:

\begin{equation}
{\cal S}_{\rm att} = \exp(- \sigma^2_{\rm nt}) \:\:\:\: {\rm with} 
\:\:\:\: \sigma^2_{\rm nt} \propto \frac{1}{N_{\rm ph}}
\label{nt}
\end{equation}
where ${N_{\rm ph}}$ is the number of photons detected by the wavefront
sensor. $\sigma^2_{\rm nt}$ includes both a noise component and a
servolag error component. Indeed thanks to the modal control, and more
precisely to the optimization of the mode gains, it is the global
contribution of noise plus servolag error which is minimized. However,
this global error scales in the same manner as the pure noise error,
which justifies the form of the function we chose in the above
equation. From WFS data, we have been able to derive -- independently
from the determination done above from the image Strehl ratio -- the
variance of the noise + servolag errors. Typical values are 0.1 rd$^2$
at 500 nm for a magnitude $R_{GS}=9.5$, 1 rd$^2$ for $R_{GS}=11.2$,
and 10 rd$^2$ for $R_{GS}=15.5$, values which are in good agreement
with the Strehl loss derived from the infrared images.
The function in Eq.~\ref{nt} was then extrapolated to other wavelengths using
a dependence $\sigma^2_{\rm nt} \propto \lambda^{-2}$. The magnitude
for which the Strehl ratio attenuation is 50\% is $R_{GS}$ = 15.7 for
$K$ band images, 15.0 at $H$, and 14.4 at $J$.  A direct extrapolation
to the $R$ bandpass -- less meaningful at this wavelength where low
Strehl ratios are usually obtained -- gives $R_{GS}$ = 13.0. We do not
list this as being the limiting magnitude of the system, for which we
know no satisfactory or unambiguous definition. For actual projects,
the limit depends more on the scientific goal one wants to achieve,
coupled with the turbulence conditions encountered at the time of
observation. As an illustration, we have achieved FWHM = 0\farcs17 in
the $K$ band using a guide star with $R_{GS}$ = 17 under good
conditions (0\farcs38 seeing).

In real observing situations, from the knowledge of $r_0$, one can
predict the performance (Strehl and FWHM) at any wavelength and any
guide star magnitude using  Figures 5 to 9.

To conclude this section, Table 2 summarizes the
Strehl ratio and FWHM expected of compensated images at various
wavelengths for median seeing conditions. ${\cal S}_{\rm static}$ is
the Strehl ratio of images delivered by the optics of PUEO and the
camera at the various wavelengths. As discussed above, the Strehl
ratios in line 6 of the table were corrected for these static
aberrations.  The last two lines give the gains in Strehl ratio and
FWHM over the uncompensated case, as defined above.

\section{IMAGE PROPERTIES}
\label{image_properties}

In this section, we try to derive global properties of the compensated
images. The boundary between the discussion in this section and that
in the previous section is not always clear. The aim is to make
this section more general to adaptive optics compensation, versus the
previous one where the actual PUEO system results were presented.

The partial correction image profile, with a coherent core -- broadened
by tip-tilt residuals -- on top of a diffuse halo, is well known.  This
partial correction profile is seen in all our near infrared images. In
the visible regime, the images are shaped more like uncompensated
images, with a profile resembling that of a Lorentzian.  In both cases,
the PSFs turned out to be very stable, and PSF fitting algorithms such
as those used in DAOPHOT have been successfully applied with excellent results, even
in extremely crowded fields at the Galactic Centre and M31 (Davidge et
al. 1997a, b).  Many examples of partial correction profiles can be found in
the literature, hence we did not feel necessary to report such a
profile here. Instead, we will go one step further and explain why
the PSF has such a shape.

A very educational and global way to understand the effect of the
compensation by an AO system is to consider the phase itself. The
phase structure function $D_{\varphi}$ is a powerful tool to
investigate the phase properties and image characteristics.  With some
limitations, it is possible to derive the phase structure function
from the point spread function. In this work, we used the following
estimator:
\begin{equation}
D_{\varphi} = -2 \times \log \left[\frac
{|FT({\rm image})| - noise(|FT({\rm image})|)}
{|FT({\rm psf\_stat})| - noise(|FT({\rm psf\_stat})|)} \right]
\end{equation}
averaged azimuthally. ``psf\_stat'' is the point spread function
acquired on the internal artificial source that includes all
uncorrected non-common path aberrations (mostly in the imaging
cameras). The average noise level is determined from the spatial
frequency domain lying outside the telescope cut-off frequency. The
noise on the Fourier transforms usually prevents an accurate
determination of the structure functions at separation $\ga$0.8$D$,
corresponding to spatial frequencies for which the amplitude of the
Fourier transform drops down to the noise level, i.e. close to the
cut-off frequency $D/\lambda$.  Using 18 images recorded at different
wavelengths ($J$, $H$, $H2$ -- 2.12~\mic~--~), under various seeing
conditions during two consecutive nights, it was possible to compute a
{\sl characteristic phase structure function} of the system, shown
plotted as a solid line in Figure 10. This
function is the average of the structure functions obtained from the 18
images, normalized by $D/r_0$ at the image wavelength. The error bars
were computed simply as the rms deviation of this ensemble of curves.
The structure function for a Kolmogorov type turbulence is plotted as a
dashed line for a $D/r_0=1$. The dashed--dotted line is the structure
function of the high spatial frequency component of the phase,
uncorrected by the system. We term this latter function the {\sl Noll
structure function}, by analogy with the Noll residual (Noll 1976).  It has been
computed using Monte-Carlo realizations of turbulent wavefronts, from
which the contribution of the system/mirror modes was entirely removed.
It is therefore the structure function of the compensated phase in
absence of any errors such as noise, spatial aliasing and servo-lag
error.

It is important to realize that these curves represent global
characteristics of the system. It is well known that the fluctuations
of the index of refraction are, to first order, independent of the
wavelength in the regime we are interested in (visible and near
infrared).  This means that the phase delay at any point, expressed as
a length, will be the same whatever the wavelength.  This is what
allows us to plot the characteristic (in the sense that it is
achromatic) structure function of the turbulent wavefront for a given
$D/r_0$.  Because the wavefront compensation by an adaptive optics
system is also an achromatic process, in the sense that the system
compensates by inducing a single phase delay at any point using a
reflective element, the properties of the compensated phase will also
be independent of the wavelength. The wavelength dependency comes at
the image formation level.

Several remarks can be made about Figure 10: 
\begin{itemize}
\item Unlike the uncompensated structure function, the Noll
structure function saturates and forms a ``plateau'' over most of the
separation domain. This ensures partial coherence over the whole
telescope pupil. Depending on the height of the plateau, this will
allow the formation of a coherent core in the image.  Saturation takes
place for separations $\rho$ for which the $\varphi(\vec{r})$ and
$\varphi(\vec{r}+\rho)$ becomes {\sl uncorrelated}.
Therefore the phase structure function at these separations can be
expressed as
\begin{equation}
D_{\varphi} = < | \varphi(\vec{r}) - \varphi(\vec{r} + \vec{\rho}) | ^2 > = 
2 < \varphi(\vec{r}) ^2 > = 2 \sigma_{\varphi}^2
\end{equation}
In other words, the structure function saturation value is twice the
phase variance over the pupil. In addition, we know that ${\cal S}
\approx \exp(- \sigma_{\varphi}^{2})$.  Consequently, the higher the
``plateau'', the smaller the coherence and the smaller the Strehl ratio
(assuming that the coherent energy can be
identified with the Strehl ratio).
\item In the Noll structure function, the saturation takes place at a
separation of approximately 70 cm, which is roughly the
``inter-actuator'' distance $d$. This is not a coincidence: naturally,
only phase corrugations of scale larger than the distance between two
actuators can be corrected.
\item The {\sl phase structure function achieved by the system} (solid
line) is larger than the Noll structure function at all separations.
The ``plateau'' is partially destroyed. The rising function at scales
larger than the inter--actuator distance can only be explained by the
presence of low--spatial frequency aberrations, such as tip--tilt,
defocus, etc., which have not been fully corrected by the system.  A
more detailed analysis shows that, when using bright guide stars, these
low order modes are principally the result of spatial aliasing (noise
and servo-lag errors are small because of the large number of photons
available for these particular examples and because the system
bandwidth -- 70 to 100 Hz -- is several times larger than the Greenwood
frequency). Overall, both our numerical simulations and experimental
results show that for curvature systems,  spatial aliasing induces a
phase error which is comparable in amplitude to the undermodeling
error. As for impact on the resulting images, the difference between
the Noll structure function and the system structure function means a
smaller Strehl ratio, and a not--fully diffraction--limited core, but
one slightly larger than the Airy pattern.
\item The decorrelation at small scales in the system structure
function is of the same type (although different quantitatively, see
below) than the loss of coherence induced by seeing. This is what
forms the compensated image ``halo''.  However, the system structure
function increases more slowly than the uncompensated one at small
scales. The {\sl cause} of this lies in the fact that large scale
perturbations play a non negligible role in the phase decorrelation at
small scales (especially because the power in these large scale
perturbations is so large), therefore, correction of perturbations of
scales $> d$ affect the phase structure function at scales $< d$.  The
{\sl consequence} of this is that the coherence of the wavefront is
increased everywhere, and in particular at small scales. The net effect
for the behavior of the structure function at small scales is
equivalent to having a larger $r_0$ value.  The ``halo'', directly
linked to the wavefront coherence at small scale, will therefore have
properties different from normal long exposure images. Because the
saturation breaks the scaling invariance of the structure function, the
characteristics of this halo depends on the wavelength of observation.
Qualitatively, at short wavelengths where the correction is only
partial, this halo will be narrower than the uncompensated seeing image
(because the coherence length is increased). At longer wavelengths, for
which good corrections are obtained, this phase decorrelation at small
scale acts like ground glass and the halo FWHM grows as $\lambda$.
A more quantitative explanation of this phenomenon will be reported
elsewhere. 
\end{itemize}

As shown above, all the observed properties of the AO--compensated
images can be explained by consideration of the phase structure
function. In addition, the phase structure function expresses the
correlation of the phase independently from the image formation
process. {\sl Knowledge of this function characterizes entirely the
performance of a given system} (in the bright guide star regime) and
allows computation of the image at {\sl any} wavelength.

Figure 11 shows the image Strehl ratio for infrared
images versus the normalized FWHM, as defined earlier. In this figure,
the crosses refer to guide stars brighter than ${\rm m}_{\rm R} = 13.5$
and the filled circles to guide stars fainter than ${\rm m}_{\rm R} =
13.5$.  These curves show that for bright guide stars, there is a {\sl
very} tight relation between the normalized FWHM and the Strehl ratio,
as noted by Tessier (1995).  For any given Strehl ratio there is a
well--defined normalized FWHM and therefore image shape, {\sl whatever
the wavelength and the atmospheric conditions}.  This is expected if
one considers the interpretation of  AO compensation in terms of the
structure function as discussed above. This link between Strehl ratio
and FWHM is however difficult to exploit when it comes to determining
the Strehl from the FWHM, especially in the domain of high Strehl
${\cal S} > 0.3$ where small variations in FWHM correspond to large
Strehl variations.  Until a more detailed study is carried out
to narrow down the
Strehl--FWHM relationship, this is at
best as a crude indicator of Strehl. Moreover, for higher order
systems, the ratio between the diffraction limited core width and the
halo width will be larger, and therefore the Strehl--FWHM relationship
will be even more bi-modal than the one reported in
figure 11, rendering any Strehl determination with this
method difficult, or at least very noise sensitive.

Use of faint guide stars modifies the relationship between the Strehl
ratio and the normalized FWHM. In terms of modal decomposition, this can
be easily understood if one considers that the noise propagation on the
correction modes behaves differently than other sources of error.
For instance, it is known that noise propagates in a very large part
on tip-tilt (especially for curvature systems), and therefore the
error on tip-tilt, {\sl relative to that of other modes}, will be
larger when dominated by noise measurement error, broadening the image
further, and modifying the Strehl/FWHM relation.
Because of insufficient data, we have not been able to investigate how
this Strehl--FWHM varies when anisoplanatism comes into play.

\section{SCIENTIFIC PROGRAMS}
\label{scientific_programs}

The exploitation of PUEO for regular scientific programs began
in August 1996.  It already includes surveys of young stars and multiple stars
in the Pleiades (Bouvier et al. 1997), imaging of comet Hale-Bopp,
the nucleus of M31 (Davidge et al. 1997a), the Galactic center (Davidge
et al. 1997b), Seyfert galaxies, QSO host galaxies (Hutchings et al. 1997) and high redshift
galaxies. Figure 12 shows the result of a 15 minute observation of
the center of our Galaxy in the $K$ band (Rigaut et al. 1997). A ${\rm m}_{\rm R} = 14.5$ star
approximately 20 arcsec away from the center of the image was used for
wavefront sensing. The galactic center was between 1.7 and 2 airmasses
during these observations, yet the uncompensated seeing was between
0\farcs65 and 0\farcs7. 
In this 13$\times$13 arcsec square image, we have detected
more than a thousand stars with a completeness limit of $K\sim$ 16. The faintest
objects have  $K\sim19$. The detection is mostly
limited by the crowding of the field but, even so, 25 stars per square
arcsecond are visible on the original image. Similar, spectacular
images on other targets amply demonstrate the potential of adaptive
optics for astronomical observations.

\section{SUMMARY}

The characteristics and performance of the CFHT adaptive optics
bonnette have been obtained from an extensive data set taken at
wavebands from $B$ to $K$ during 16 nights. These data give many
results on both the typical atmospheric conditions encountered on
Mauna Kea and on the adaptive compensation. These include:

\begin{itemize}
\item Most of the time (80-85\%) the
observed atmospheric turbulence was well characterized by Kolmogorov turbulence.
\item The median value of $r_0$ at 500nm was 15.5cm, corresponding to a
zenith-corrected seeing disk of FWHM = 0\farcs58.
\item Analysis of the variation of the observed Strehl ratios as a function
of $r_0$ indicates that the wavefront correction achieved by PUEO corresponds to
full compensation of eight Zernike modes.
\item The improvement in the Strehl ratio peaks at $r_0 \sim$ 50cm (J
band in median seeing conditions). The peak intensity of images is
improved by a factor twelve at J band and by more than a factor five from $I$ to $K$.
\item Images with FWHM$\sim$0\farcs1 are common from $I$ to $K$, and are
basically diffraction-limited at $H$ and $K$. Significant ($>$2) gains
in FWHM are realized at all wavelengths from $B$ to $K$. The maximum gain
in FWHM occurs for $r_0$ $\sim$ 40 cm.
\item A decrease in performance becomes apparent for guide stars
fainter than $R$ = 13.5. Strehl ratios are reduced by 50\% at $R$ = 15.7
in $K$, 15.0 at $H$, 14.4 at $J$ and $\sim$13 at $R$ under median seeing
conditions.
\item The observed properties of the PSFs of the compensated images
are very stable. We have been able to derive the phase structure
function achieved by our system, and proposed an interpretation of
this function that explains the global properties of AO compensated
images.  The phase structure function achieved by the system is, as
expected, larger than if only fitting error is considered, largely
because of spatial aliasing. One result of this is that the cores of
the images are not fully diffraction--limited. The haloes of the
images are also affected, but in a very different way, leading to
suppressed wings and a lorentzian profile at short wavelengths.
\end{itemize}

We believe that the operation and the results of this instrument prove
that adaptive optics is now a mature technique and can be applied
efficiently -- in terms of angular resolution, but also in terms of
overhead and sensitivity -- to astronomy.  Our very simple user
interface was operated during several nights by astronomers with no
{\it a priori} knowledge of adaptive optics and proved to be very
efficient, robust and reliable. The overhead involved in acquiring
astronomical data is comparable to, or even shorter than, that of
regular imaging devices since the system automatically achieves
perfect guiding and focus. An interactive ``AOB performance meter''
at http://www.cfht.hawaii.edu/manuals/aob/psf.html provides an estimate of the expected image quality under given
seeing conditions for various brightnesses and distances of the
guide star. Thanks to the excellent natural seeing at
Mauna kea, we can now expect diffraction--limited performance in terms
of FWHM, for guide stars as faint as $R$ = 16, offering 0\farcs1
resolution to many fields in astronomy, including extragalatic
studies.

\acknowledgements
We would like to thank all the people involved in the development and
operation of the AOB at CFHT, DAO, Observatoire de Meudon and CILAS. In
particular, John Kerr, Jerry Sovka, Scot McArthur, Greg Barrick, Barney
Magrath, Grant Matsushige, Daniel McKenna, Daniel Sabin, Jerome
Bouvier, Bill Cruise, Claude Berthoud, Suzan Wood, Linda Fisher, Guy
Monnet and Pierre Couturier at CFHT; Dennis Derdall, Walter Grundman,
Jim Jennings, Brian Leckie, Rick Murowinski, Allan Moore, Scott
Roberts, Les Saddlemyer, Jerry Sebesta, and Bob Wooff at DAO; Sen Wang,
Claude Marlot and Alain Piacentino at the Observatoire de Meudon;
Jean-Paul Gaffard, Jacques Peysson, Jean-Jacques Roland, Gerard Zeins
and Patrick Petitgas at CILAS.  We are indebted to Francois Roddier,
J.Elton Graves and Malcolm Northcott at the Institute for Astrophysics,
University of Hawaii, for valuable advice on curvature systems.
Special thanks to Ren\'e Doyon and Daniel Nadeau of Universit\'e de
Montr\'eal for modifying, lending and operating their MONICA infrared
camera during all three runs.

\clearpage

\begin{deluxetable}{lll}
\tablewidth{43pc}
\tablenum{1}
\tablecaption{Characteristics of the CFHT Adaptive Optics Bonnette}
\tablehead{
\colhead{Subassembly}   & \colhead{Characteristic}           &  \colhead{Description}
}
\startdata
{\bf Optomechanics} &
Total number of mirrors in science train & 5 + 1 beamsplitter (transmission) \\
& Total number of mirrors in WFS train 	& 9 + 1 beamsplitter (reflection) \\
& Transmission of Science train		& 70\% (V) excluding beamsplitter \\
&					& 75\% (H), 70\% (K) including dichroic \\
& Input/Output focal ratios		& f/8 , f/19.6 \\
& Overall Bonnette dimension		& Diameter 120 cm, Thickness 28 cm \\
& Flexure				& $<$10 \mic per hour between WFS\\
&                                       & and science focus \\
& Optical quality			& $\lambda/20$ rms at 0.5 \mic\ with DM flat \\ 
& Instrument clear field of view	& 90\arcsec\ diameter \\
& Atmospheric Dispersion Compensator    & Removable \\
&                                       & Designed for zenith distances $<$60\arcdeg \\
\hline
{\bf Wavefront sensor}
& Type					& Curvature \\
& Number of subapertures			& 19 \\
& Detectors				& APDs (45\% peak QE,\\ 
&                                       & ~~~~~~~~dark current$\sim$ 20e$^-$s$^{-1}$) \\
& Field of View				& 1-2 arcsec depending on optical gain\\
\hline
{\bf Deformable mirror}
& Type				& Curvature (+ dedicated Tip-Tilt) \\
& Number of electrodes			& 19 \\
& Stroke				& $\sim \pm$ 10 \mic\  \\
& First mechanical resonance	        & $>$ 2kHz \\
& Overall dimension			& 80 mm \\
& Pupil size on DM			& 42 mm \\
& Conjugation				& Telescope pupil \\
& Tip/tilt mirror stroke 		& $\pm$ 4 arcsec \\
\hline
{\bf Control}
& Sampling/command frequency		& Selectable (1000Hz, 500Hz, 250Hz,..) \\
& Max bandwidth 0dB rejection		& 105 Hz \\
& Max bandwidth $-$3dB closed--loop	& 275 Hz \\
& Control scheme			& Modal, 18 mirror modes controlled \\
&					& Closed--loop mode gains optimization \\ 
\hline
{\bf Instrumentation}
& CCD        & 2K$\times$2K pixels, 0\farcs03 or 0\farcs06 pixel$^{-1}$ \\
& IR imager    & NICMOS array, 0\farcs034 pixel$^{-1}$ \\
& Visible integral field spectrograph    & Undergoing commissioning

\enddata
\end{deluxetable}

\begin{deluxetable}{lllllll}
\tablewidth{33pc}
\tablenum{2}
\tablecaption{Performance Summary For Median Seeing Conditions}
\tablehead{
\colhead{Waveband} & \colhead{$V$} & \colhead{$R$} & \colhead{$I$} & \colhead{$J$} & \colhead{$H$} & \colhead{$K$} 
}
\startdata
Wavelength [\mic] & 0.54 & 0.65 & 0.83 & 1.25 & 1.65 & 2.23 \\ 
${\cal S}_{\rm static}$ & 0.50 & 0.65 & 0.75 & 0.77 & 0.84 & 0.93 \\ 
Median $r_0 (\lambda)$ [cm] & 17 & 21 & 28 & 46 & 65 & 93 \\ 
$D/r_0$                   & 21.3 & 17.1 & 12.7 & 7.8 & 5.6 & 3.9 \\ 
Strehl ratio & 0.01 & 0.02 & 0.05 & 0.21 & 0.41 & 0.61 \\ 
FWHM [arcsec] & 0.24 & 0.19 & 0.12 & 0.095 & 0.11 & 0.14 \\ 
FWHM/($\lambda/D$) & 7.6 & 5.1 & 2.5 & 1.34 & 1.18 & 1.12 \\
${\rm Gain}_{\rm Strehl}$ & 4.0 & 5.0 & 7.0 & 12.5 & 11.6 & 9.5 \\ 
${\rm Gain}_{\rm FWHM}$ & 2.6 & 3.2 & 4.8 & 5.9 & 5.0 & 3.6 \\

\enddata
\end{deluxetable}

\clearpage

\begin{figure}
\plotone{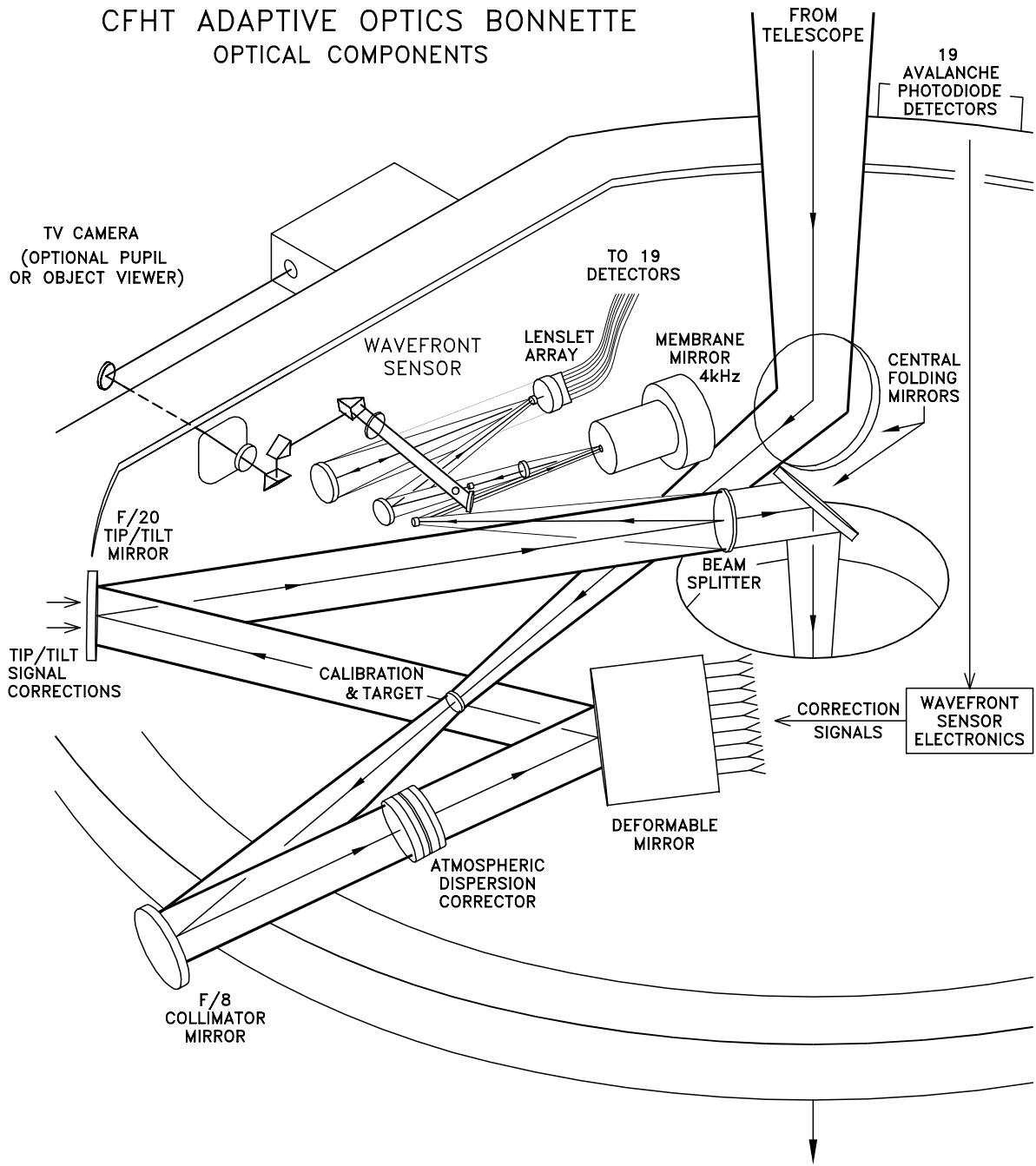}
\caption{Optical path of the instrument. The central
folding mirrors are on a movable slide, so that the direct and the
corrected focus are coincident. A calibration source and target can be
inserted at the location of the telescope focus for calibration and
testing. The wavefront sensor can be remotely moved to allow selection
of a reference star different from the science object. \label{optical_scheme}}

\end{figure}

\begin{figure}
\plotone{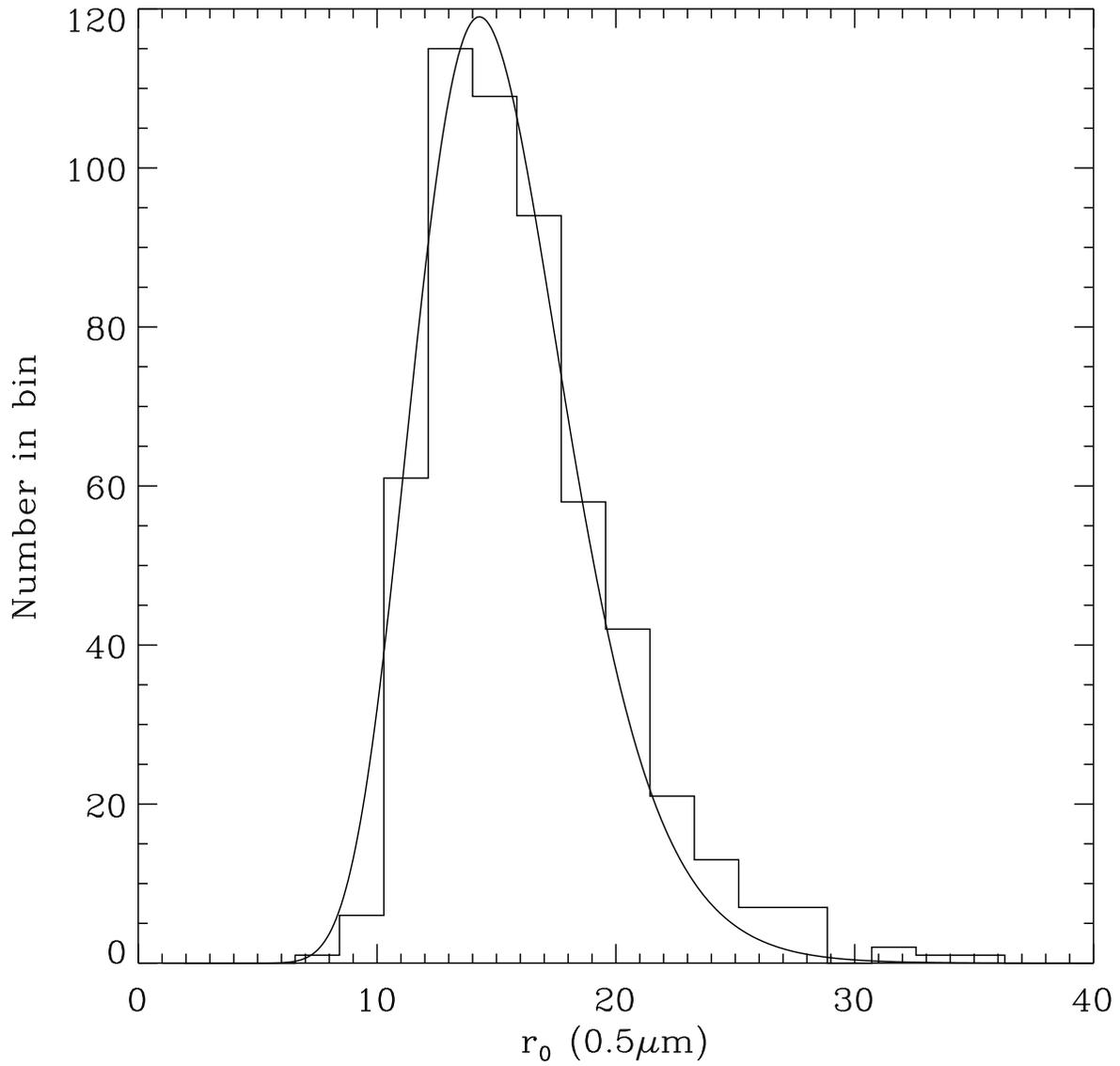}
\caption{Distribution of $r_0$ values at 500 nm at
the Cassegrain focus of the 3.6-m CFH telescope. The solid curve is a
fit using a log--normal distribution with mean equal to 15.5 cm and a
standard deviation $\sigma = 0.13$.  This median value corresponds to
an observed seeing disk of FWHM = 0\farcs67, or 0\farcs58 when
corrected to the zenith. \label{r0distribution}}

\end{figure}

\begin{figure}
\plotone{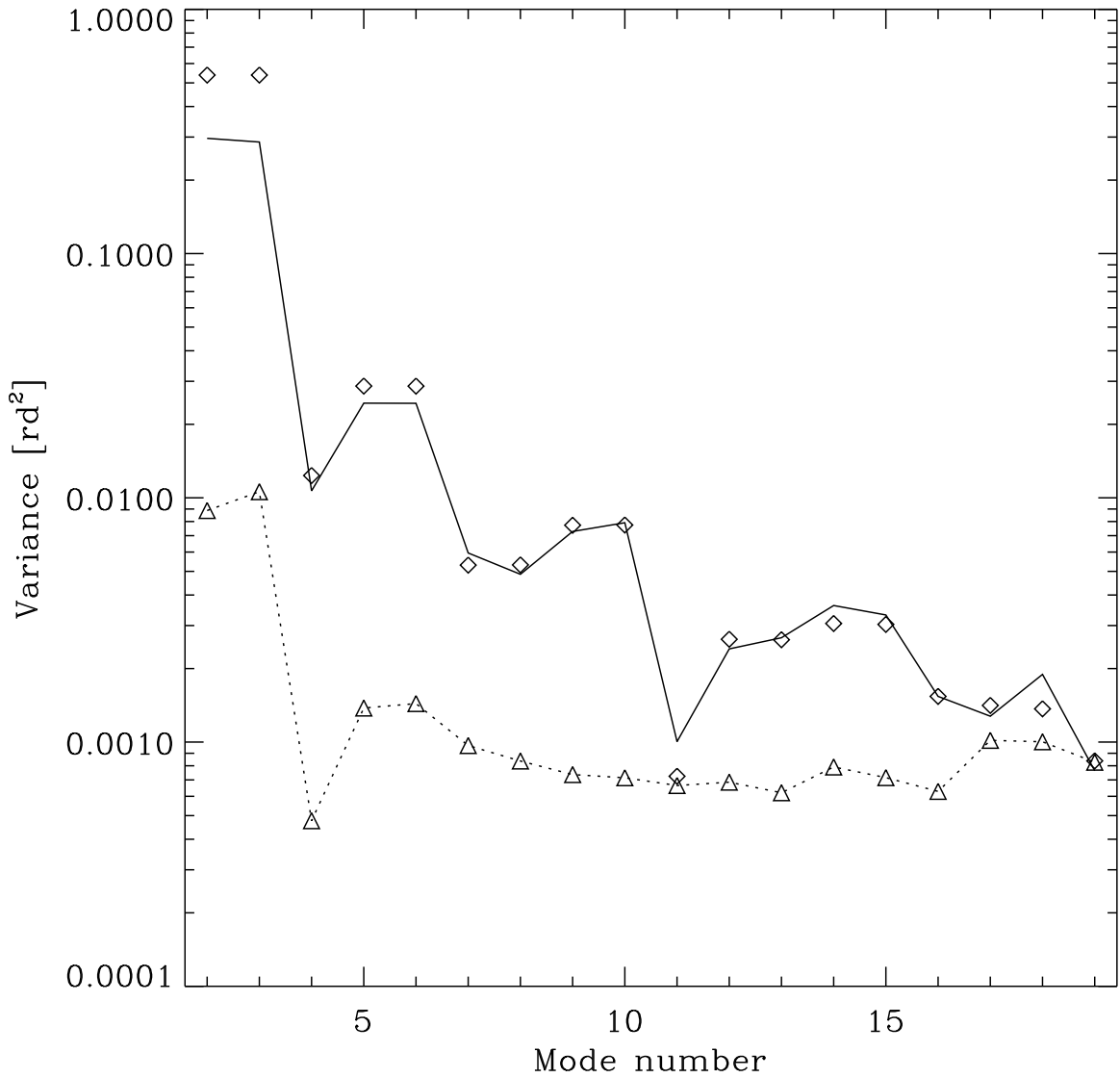}
\caption{An example of the mirror mode coefficient
variances: input or uncompensated wavefront (solid line), Kolmogorov
fit (diamonds) and the compensated wavefront (triangles and dashed
line). The data were normalized for $D/r_0=1$ and corrected for noise
and spatial aliasing. \label{modevariance}}

\end{figure}

\begin{figure}
\plotone{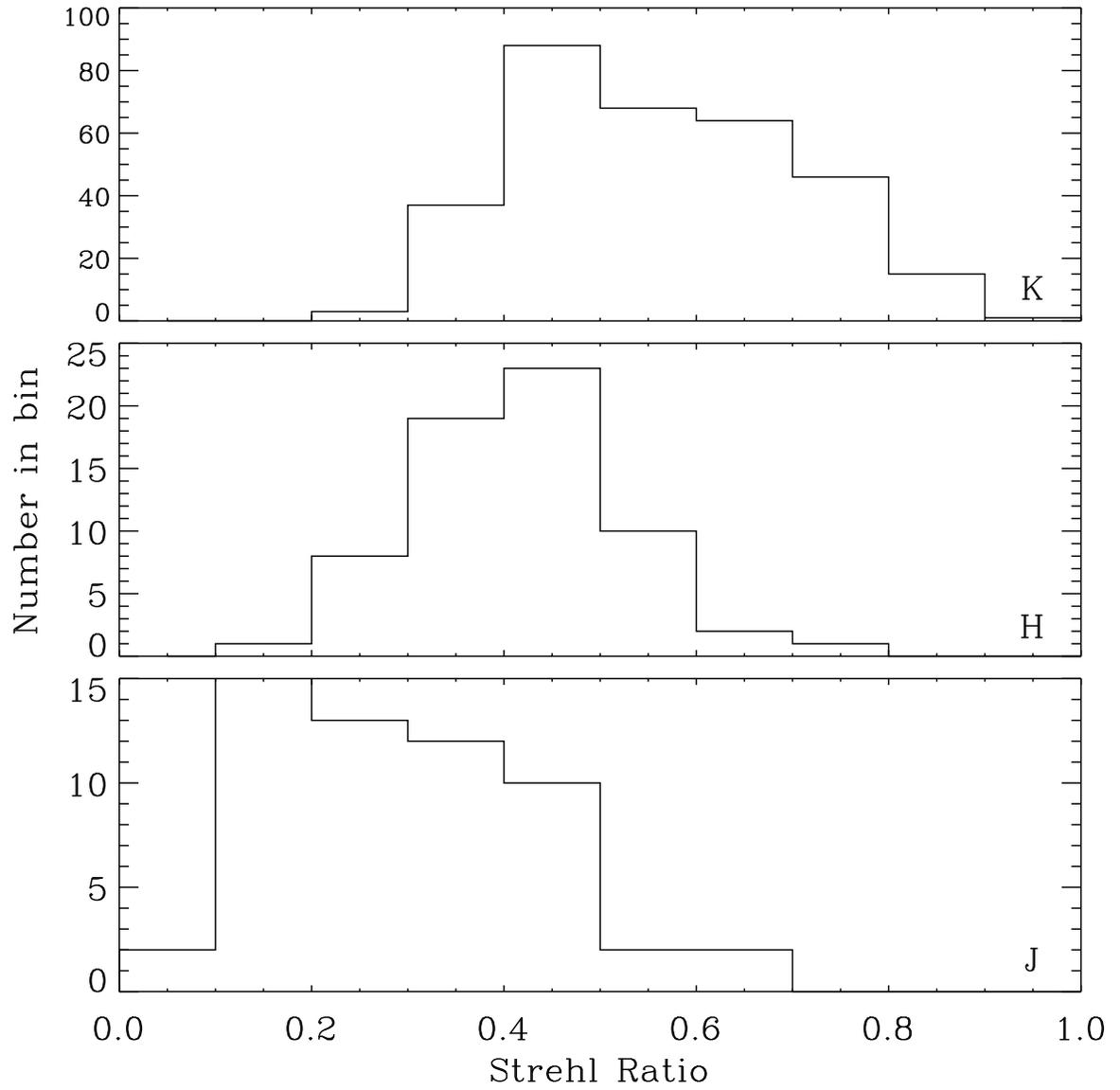}
\caption{Strehl ratio distribution for the $J$, $H$
and $K$ bands. The Strehl values are corrected for the static
aberrations of PUEO and the camera (see text). \label{strehlhist}}
\end{figure}

\begin{figure}
\plotone{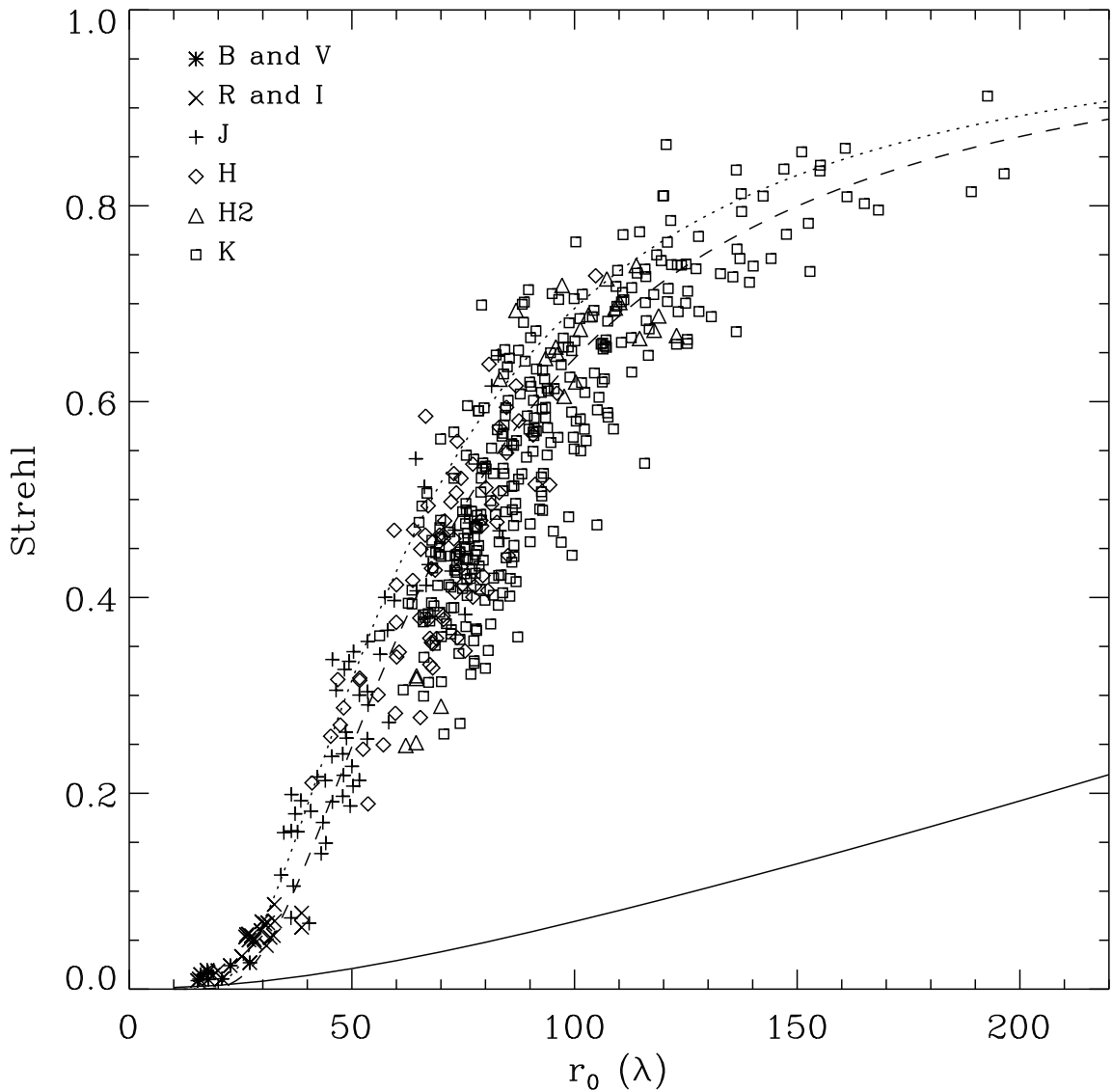}
\caption{Strehl ratio, corrected for the optical
bench static aberrations, versus the Fried parameter $r_0$ at the
image wavelength.  The bandpasses are noted with different
symbols. The lower solid line is the Strehl ratio of the
seeing--limited image. \label{strehlvsr0}}
\end{figure}

\begin{figure}
\plotone{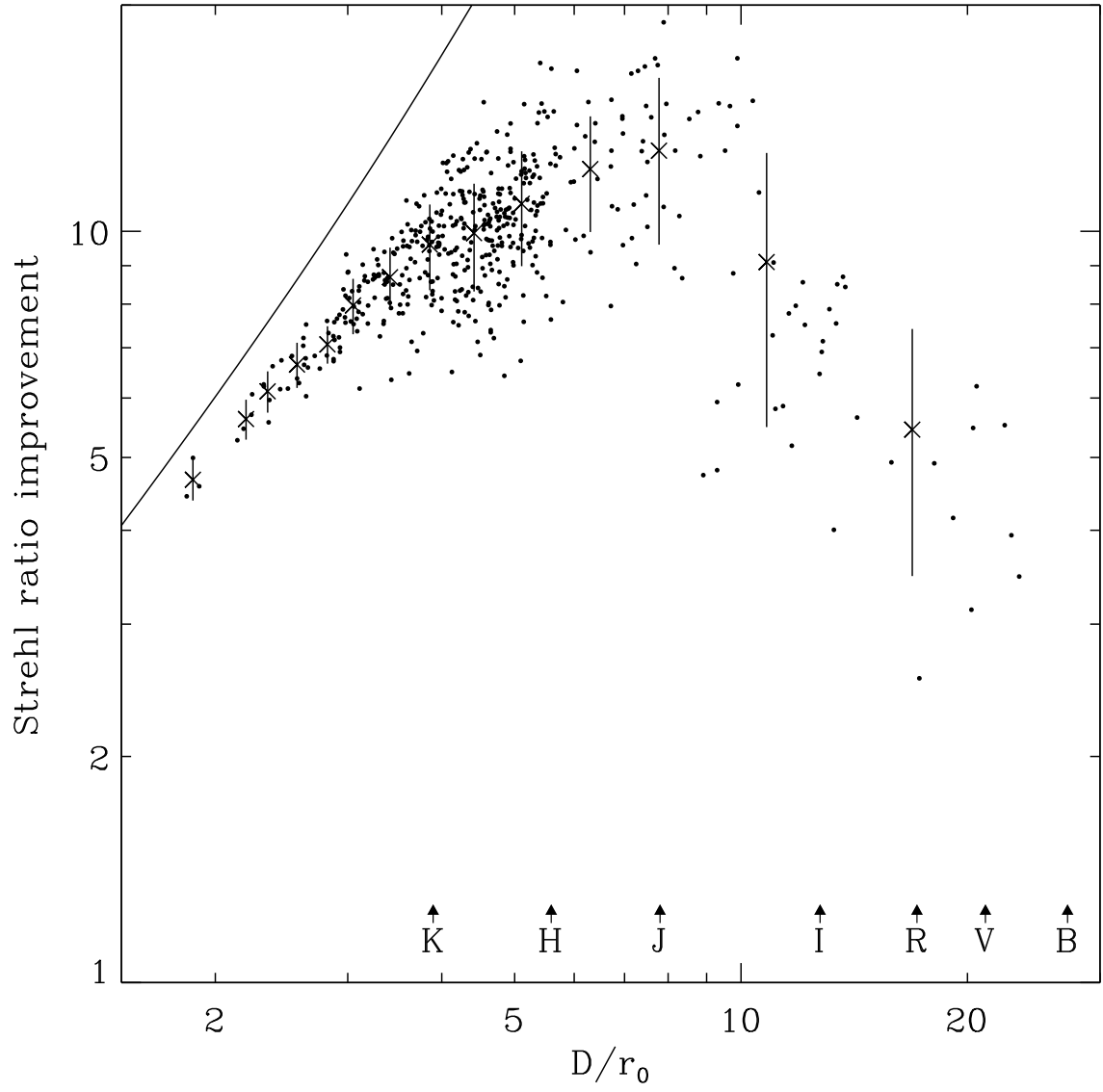}
\caption{Strehl ratio improvement versus $D/r_0$ at
the image wavelength.  In this and the subsequent two figures, $D/r_0$
values for median seeing conditions are shown at the bottom of the
figure for wavelength bands from $B$ to $K$. \label{strehlgain}}
\end{figure}

\begin{figure}
\plotone{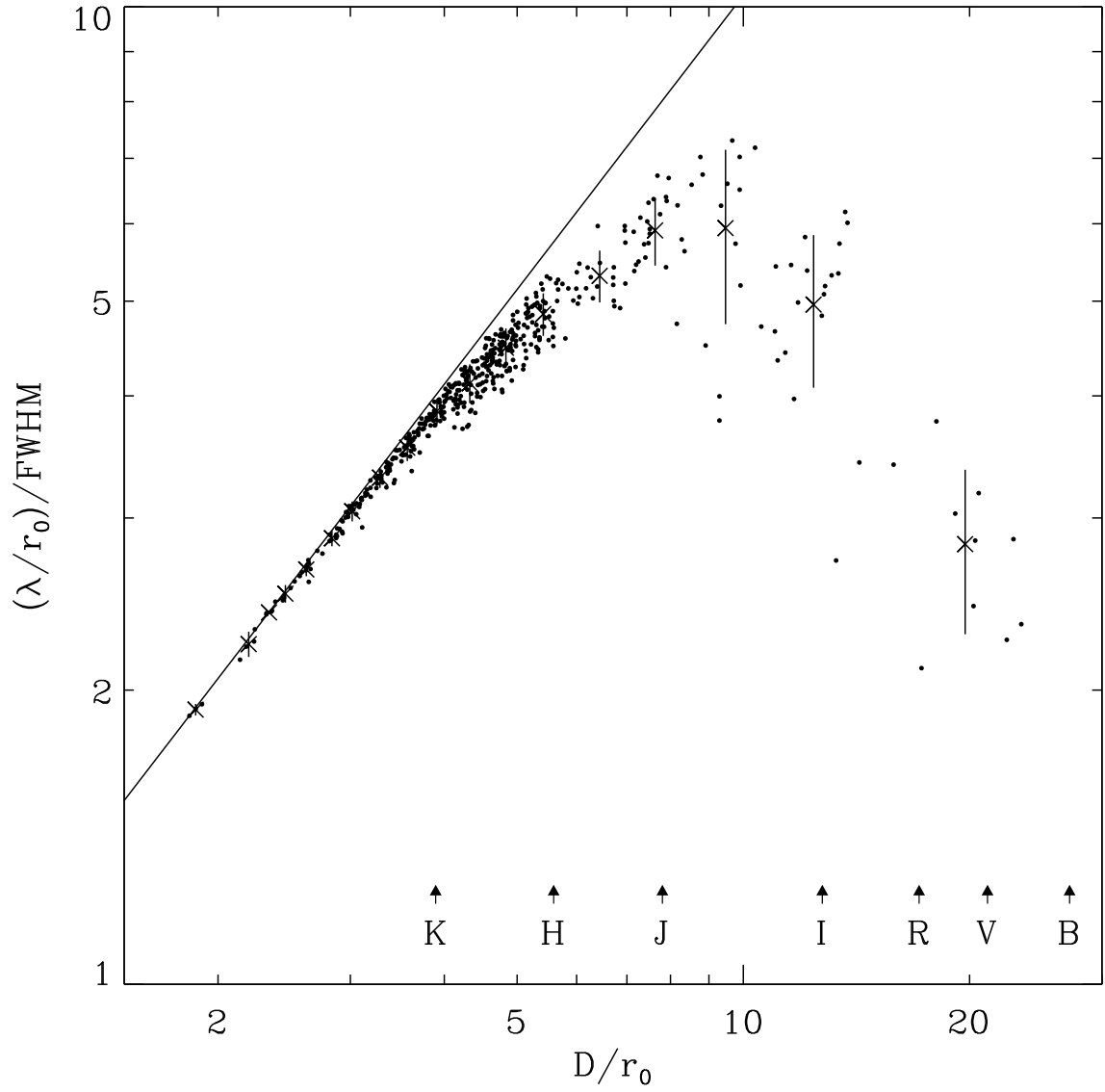}
\caption{Gain in FWHM versus $D/r_0$ at the image
wavelength. \label{fwhmgainvsr0}}
\end{figure}

\begin{figure}
\plotone{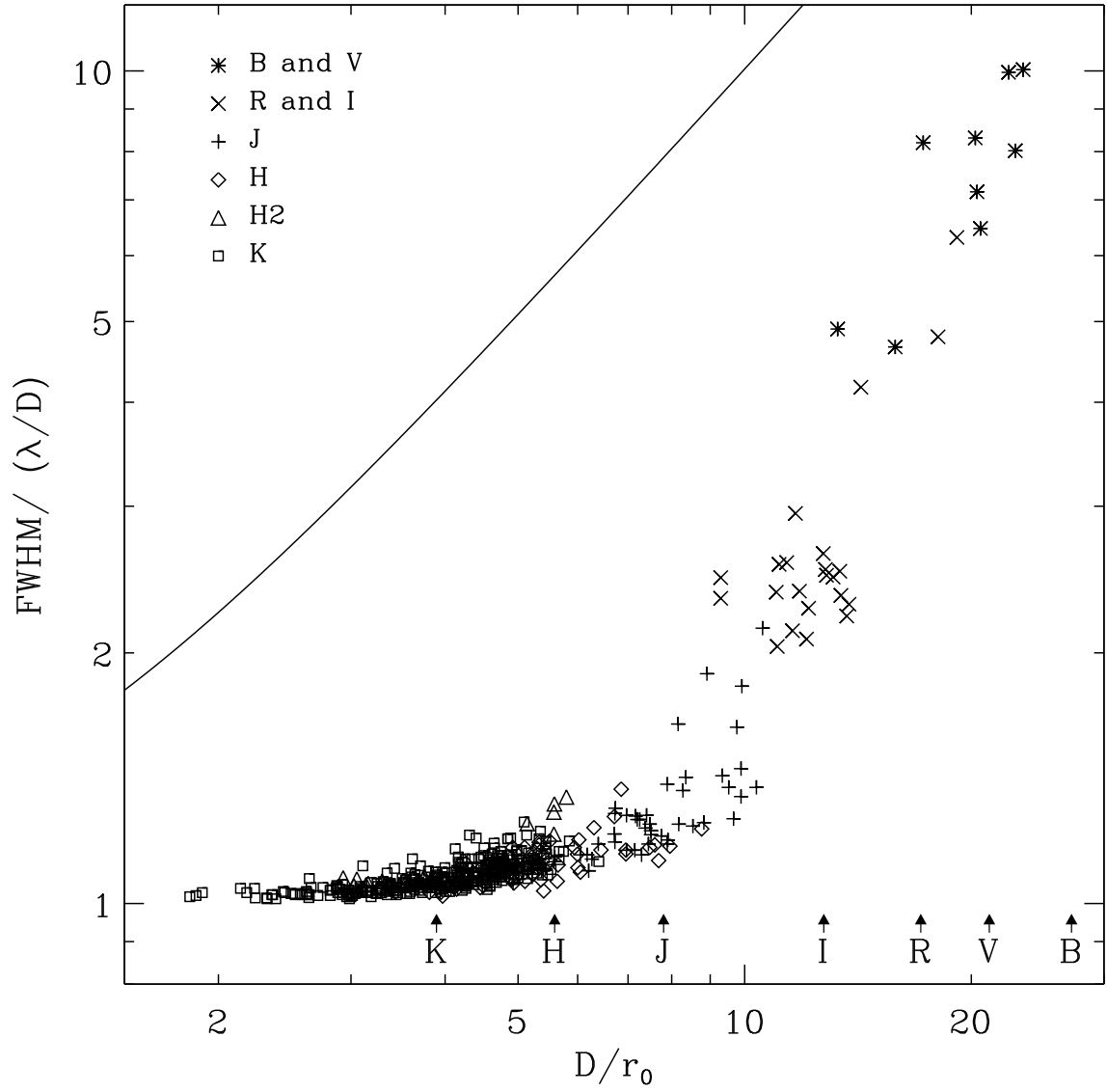}
\caption{Normalized image FWHM versus $D/r_0$ at the
image wavelength. \label{normalizedfwhmvsr0}}
\end{figure}

\begin{figure}
\plotone{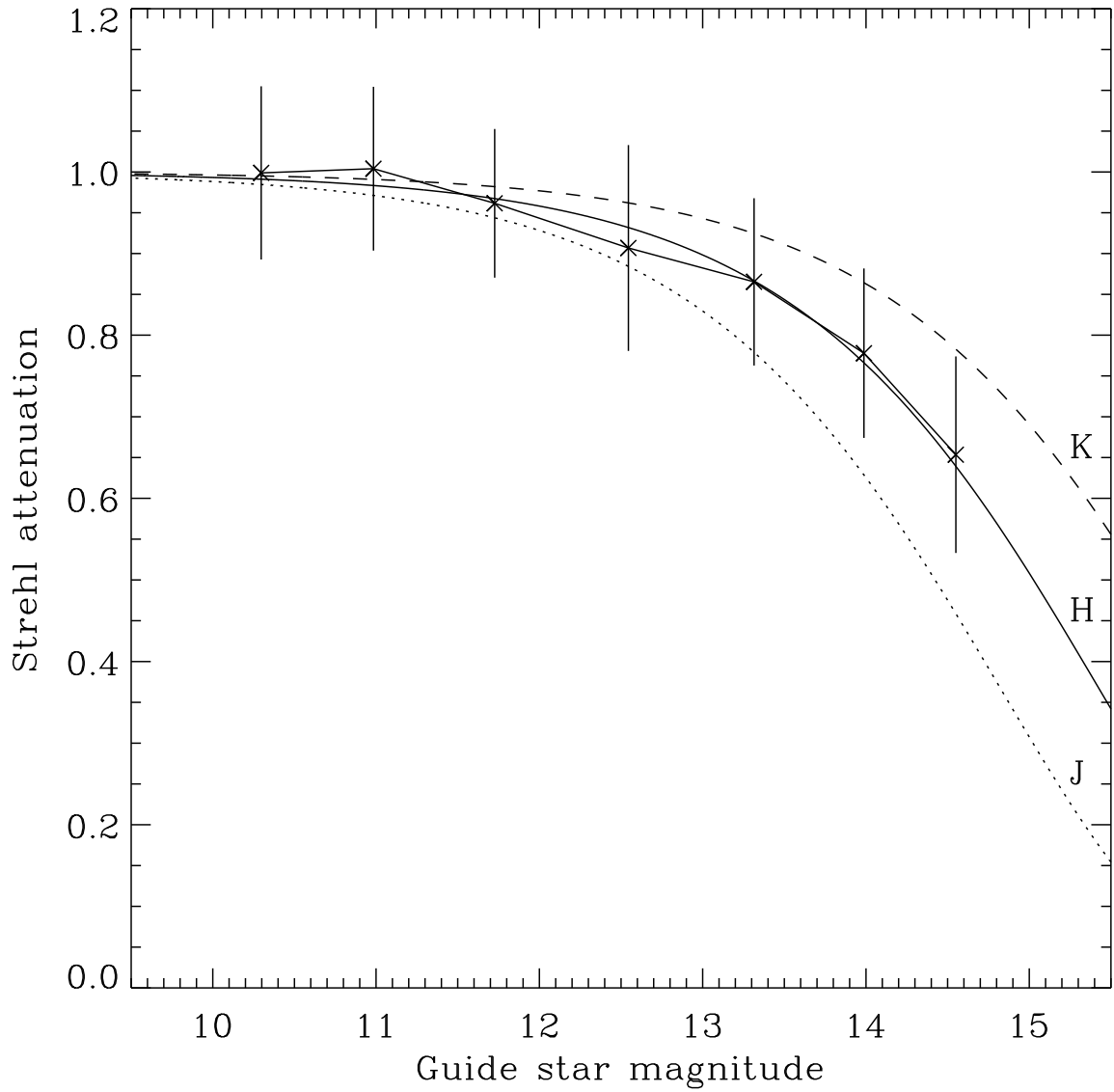}
\caption{Attenuation of the Strehl ratio compared to
the bright guide star case versus the guide star $R$ magnitude for
measurements at $H$. Average curves for $J$ and $K$ are also shown.
\label{strehlvsmag}}
\end{figure}

\begin{figure}
\plotone{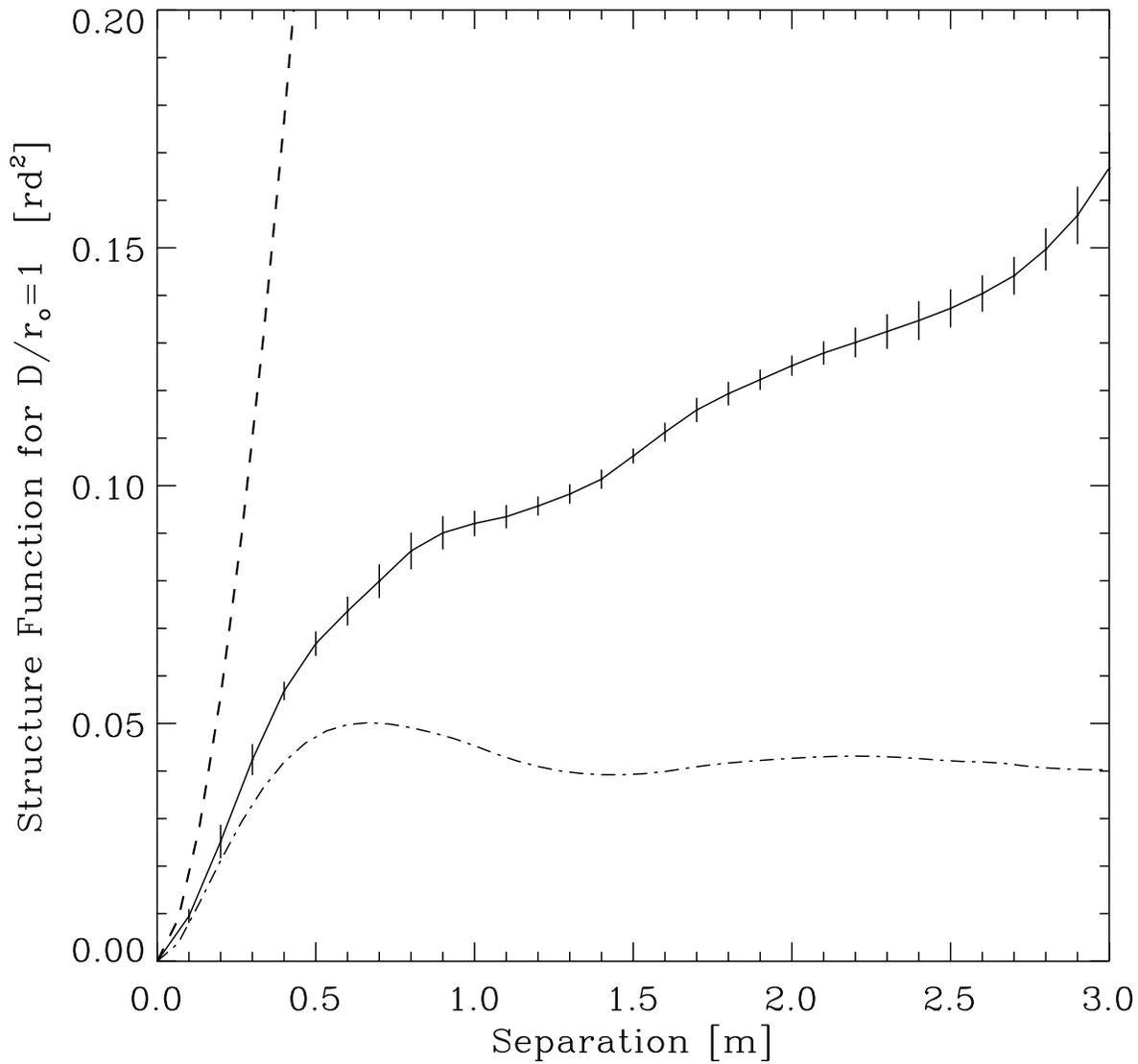}
\caption{PUEO characteristic phase structure
functions: observed (solid line + error bars), turbulent (dashed line)
and high order phase residual alone or ``Noll structure function"
(dashed-dotted line). \label{fonctionstructure}}
\end{figure}

\begin{figure}
\plotone{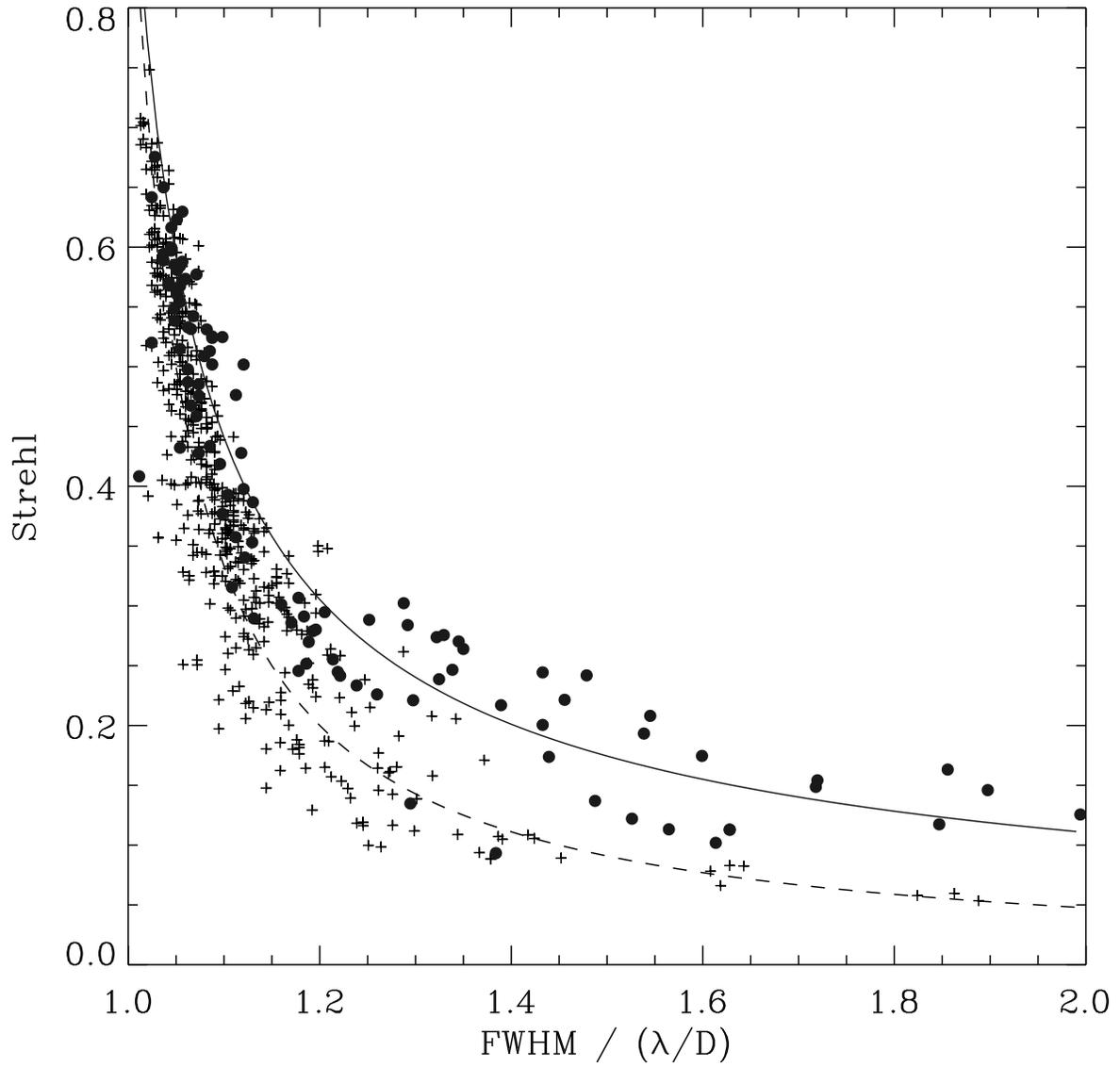}
\caption{Image Strehl ratio versus normalized FWHM
for bright stars (crosses) and faint stars (filled circles).
\label{strehlvsfwhm}}
\end{figure}

\begin{figure}
\plotone{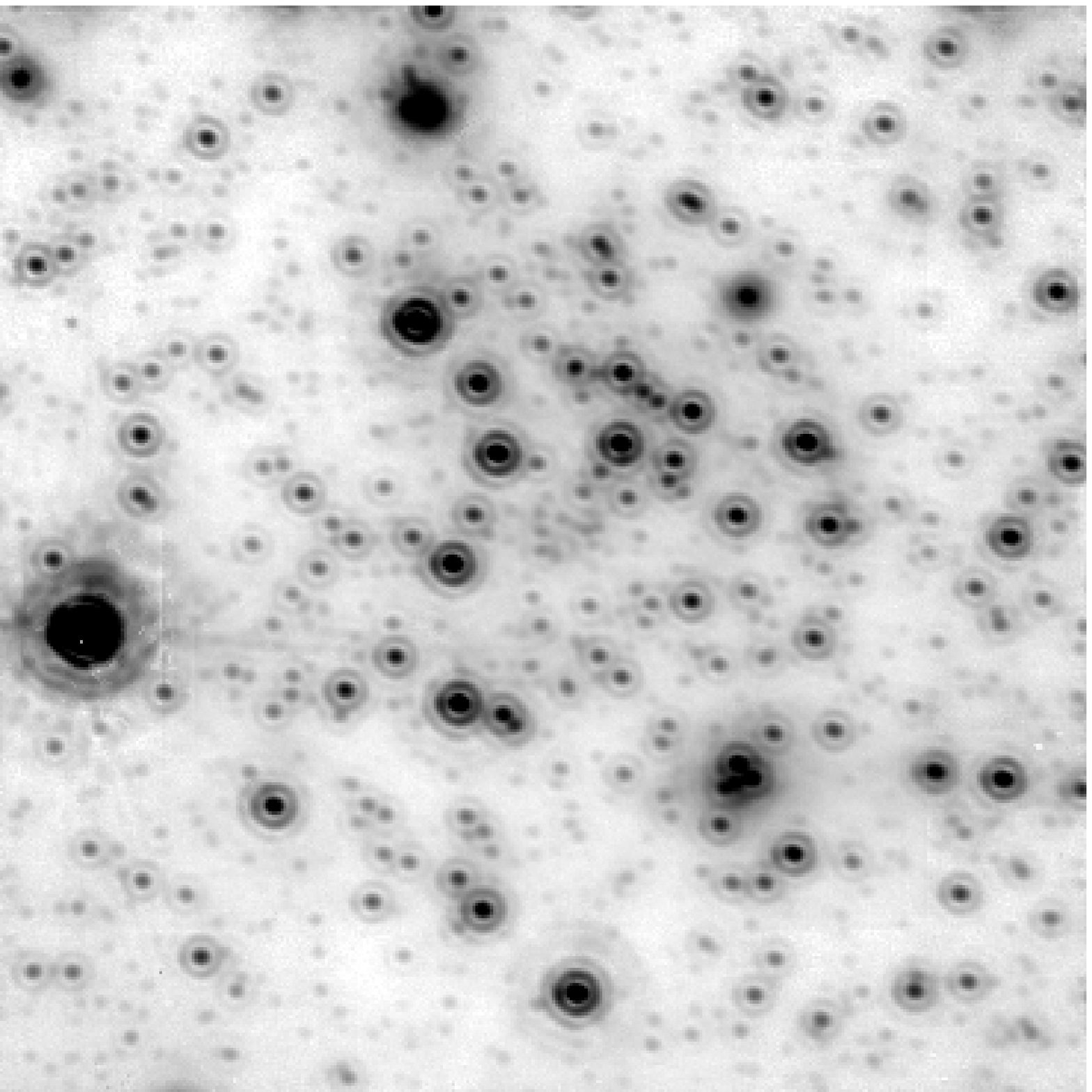}
\caption{Mosaic image of the galactic center in the
$K$ band. The field is 13\arcsec\ square and the integration time was
15 minutes. North is to the left at an angle of $-$100\fdg6 to the vertical, 
East is up at $-$10\fdg6.  The brightest objects in the field have been
saturated  and the gamma correction adjusted
on purpose to show the fainter images.  Up to 5 diffraction
rings are visible around the brightest stars on the original image.
\label{galcen}}
\end{figure}


\begin{thebibliography}{}

\bibitem{} Arsenault, R., Salmon, D.A., Kerr, J., Rigaut, F.,
Crampton, D. and Grundmann, W.A 1994, SPIE Conf, 2201, 883

\bibitem{} Beuzit, J.-L., Hubin, N., Demailly, L., Gendron E., Gigan,
P. et al. 1995, in OSA/ESO topical meeting on Adaptive Optics,
ed. M. Cullum (Garching, ESO), p 57

\bibitem{} Bouvier, J., Rigaut, F. and Nadeau, D. 1997, A\&A, 323, 139

\bibitem{} Davidge, T.J., Rigaut, F., Doyon, R. and Crampton D. 1997a,
A.J., 113, 2094

\bibitem{} Davidge, T.J., Simons, D.A., Rigaut, F., Doyon, R., Becklin, E.E. and
Crampton, D. 1997b, A.J., in press

\bibitem{} Ellerbroek, B.L., Van loan, C., Pitsianis, N.P. and
Plemmons R.J. 1994, JOSA A, Vol 11, no 11, 2871

\bibitem{} Gaffard, J.-P., Jagourel, P. and Gigan, P. 1994, SPIE Conf, 2201, 688

\bibitem{} Gendron, E. and L\'ena, P. 1994, A\&A, 291, 337

\bibitem{} Graves, J.E. and McKenna, D. 1991, SPIE Conf, 1542, 262

\bibitem{} Graves, J.E., Roddier, F.J., Northcott, M.J. and
Anuskiewicz, J. 1994, SPIE Conf, 2201, 502

\bibitem{} Hutchings, J.B., Crampton, D., Morris, S.L., \& Steinbring, E. 1997,
submitted to PASP

\bibitem{} Lai, O., Arsenault, R., Rigaut F. et al, 1995, in OSA/ESO
topical meeting on Adaptive Optics, ed. M. Cullum (Garching, ESO), p. 491

\bibitem{} Noll, R.J. 1976, JOSA, 66 (3), 207


\bibitem{} Racine, R., Salmon, D., Cowley, D. and Sovka, J. 1991, PASP,
103, 1020

\bibitem{} Richardson, E.H. 1994, in NATO ASI Series: Adaptive Optics
for Astronomy, eds. D.alloin and J.-M. Mariotti, (Carg\'ese), p.~227

\bibitem{} Rigaut, F., Rousset, G., Kern, P., Fontanella, J.-C., Gaffard,
J.-P., Merkle, F. and L\'ena, P. 1990, A\&A, 250, 280

\bibitem{} Rigaut, F., Arsenault, R., Kerr, J., Salmon, D.A., 
Northcott, M.J., Dutil, Y. and Boyer, C. 1994, SPIE Conf, 2201, 149

\bibitem{} Rigaut, F., Doyon, R., Davidge, T., Crampton, D. and Rouan,
D. 1997, submitted to Ap. J.

\bibitem{} Roddier, F.J. 1988, Appl.Opt., 27, 1223

\bibitem{} Roddier, F.J. et al. 1990, SPIE Conf, 1236, 485

\bibitem{} Roddier, F.J., Graves, J.E., McKenna, D. and 
Northcott, M.J. 1991, SPIE Conf, 1524, 248

\bibitem{} Rousset, G., Fontanella, J.-C., Kern, P. et al. 1990, A\&A, 230, L29

\bibitem{} Rousset, G. 1994, in NATO ASI Series: Adaptive Optics
for Astronomy, eds. D.alloin and J.-M. Mariotti, (Carg\'ese), p.115

\bibitem{} Tessier, E. 1995, in OSA/ESO topical meeting on 
Adaptive Optics, ed. M. Cullum (Garching, ESO), p. 465

\bibitem{} Thomas, J., Rigaut, F.J. and Arsenault R. 1997, SPIE Conf, 3126-15

\bibitem{} V\'eran, J.P., Rigaut, F.J., Rouan, D. and Maitre, H. 1997, JOSA
A, 14, 11

\bibitem{} Winker, D.M. 1991, JOSA A, Vol 8, no 10, 1568

\end{thebibliography}
\end{document}